\documentclass[9pt,twocolumn,twoside]{optica}

\setboolean{shortarticle}{false}
\setboolean{minireview}{false}

\usepackage{graphicx}
\usepackage{multirow,bigdelim}
\usepackage{bm}
\usepackage{amsmath}
\usepackage{lipsum}

\usepackage[english]{babel}
\usepackage{blindtext}

\usepackage{color}

\newcommand{\mrm}{\mathrm}
\newcommand{\Poincare}{Poincar\'{e} }

\usepackage{booktabs}
\usepackage{array}

\title{Self-similarity of optical rotation trajectories around the \Poincare sphere with application to an ultra-narrow atomic bandpass filter
}
\dates{\today}

\author[1,*]{James Keaveney}
\author[1]{Dennis A. Rimmer}
\author[1]{Ifan G. Hughes}

\affil[1]{Joint Quantum Centre (JQC) Durham-Newcastle, Department of Physics, Durham University, South Road, Durham, DH1 3LE, United Kingdom}
\affil[*]{james.keaveney@durham.ac.uk}

\begin{abstract}
We present an investigation of magneto-optic rotation in both the Faraday and Voigt geometries. We show that more physical insight can be gained in a comparison of the Faraday and Voigt effects by visualising optical rotation trajectories on the \Poincare sphere. This insight is applied to design and experimentally demonstrate an improved ultra-narrow optical bandpass filter based on combining optical rotation from two cascaded cells - one in the Faraday geometry and one in the Voigt geometry. Our optical filter has an equivalent noise bandwidth of 0.56~GHz, and a figure-of-merit value of 1.22(2)~GHz$^{-1}$ which is higher than any previously demonstrated filter on the Rb D2 line.
\end{abstract}

\setboolean{displaycopyright}{false}

\begin{document}

\maketitle


\section{Introduction}

The quantitative understanding of atom-light interactions in atomic vapours is a key component in developing new applications based on atomic and optical physics. The study of linear and non-linear magneto-optical effects in atomic media has been particularly fruitful; a review of the field can be found in ref.~\cite{Budker2002}. Applications include magnetometry for medical imaging~\cite{Johnson2013,Marmugi2016,Alem2017} and remote sensing~\cite{Lee2006}, atom-based optical isolators~\cite{Weller2012b}, methods for laser frequency stabilisation far from resonance~\cite{Zentile2014a,Reed2018}, and optical bandstop~\cite{Miles2001,Rudolf2014,Uhland2015}, dichroic~\cite{Abel2009a}, and bandpass~\cite{Ohman1956b} filters.
Of these applications, atomic bandpass filters continue to generate much attention across a wide range of research areas, from astronomy~\cite{Hart2016} to oceanography~\cite{Rudolf2014}.

First conceived in the 1950s~\cite{Ohman1956b}, optical bandpass filters based on the Faraday effect were investigated (and occsaionally re-invented) throughout the 1970s~\cite{Agnelli1975,Cacciani1978} and into the 1980s~\cite{Yeh1982}. 
They underwent a renaissance in the 1990s~\cite{Dick1991,Wanninger1992,Menders1992,Yin1992,Chen1993,Chan1993,Hu1993} which was also when the FADOF acronym (Faraday Anomalous Dispersion Optical Filter) was coined. 
In the present day, their use has extended to hybrid systems integrating quantum dots with atomic media~\cite{Portalupi2016}, 
atmospheric LIDAR~\cite{Fricke-Begemann2002},
quantum key distribution~\cite{Shan2006},
temperature profile measurements of the ocean~\cite{Rudolf2012,Ling2014} and
self-stabilising laser systems~\cite{Miao2011, Keaveney2016c}. 
%
%
\begin{figure}[t]
\begin{center}
\includegraphics[width=0.8\columnwidth]{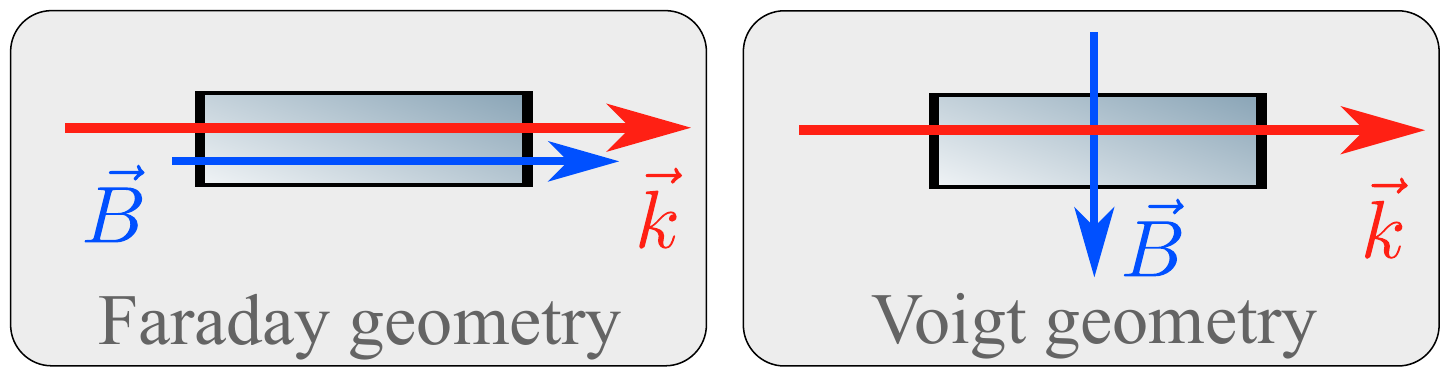}
\caption{Illustration of the Faraday and Voigt geometries, showing the relative orientation of the wavevector $\vec{k}$ and the magnetic field $\vec{B}$.
}
\label{fig:geometry}
\end{center}
\end{figure}
Though most filters in the literature use the Faraday effect, there have been some filters based on the Voigt effect~\cite{Kessler1965,Menders1992,Yin2016}.
In addition, there is now interest in a more general type of atomic filter that does not necessarily use the Faraday or Voigt effects~\cite{Rotondaro2015,Keaveney2018a} to improve filter performance with a slight modification to the experimental arrangement.
%

%
\begin{figure*}[t]
\begin{center}
\includegraphics[width=1.9\columnwidth]{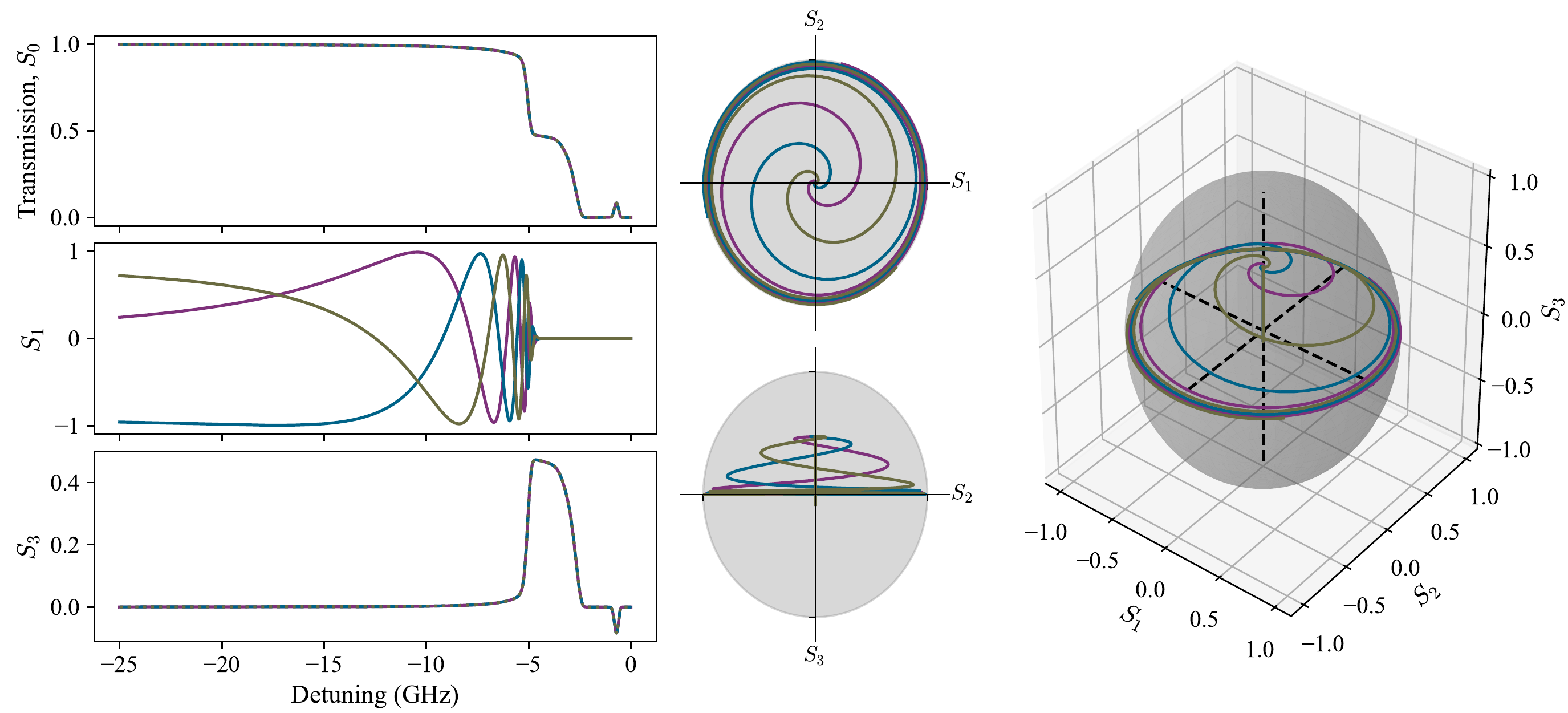}
\caption{Faraday rotation around the Rb D2 resonance lines visualised on the \Poincare sphere. The spectra show the Stokes parameters $S_0$, $S_1$ and $S_3$. In the Faraday geometry the polarisation rotates around the $S_3$ axis; trajectories with different initial linear polarization angles ($\pi/4$, purple; $\pi/4+\pi/3$, blue, $\pi/4-\pi/3$, olive) are self-similar - i.e. when viewed in the plane perpendicular to the rotation axis, they follow the same trajectory but are rotated with respect to each other. Far from resonance the polarisation vector starts at the equator of the \Poincare sphere. As the frequency moves towards resonance, the vector rotates around the equator, while still maintaining linear polarisation ($S_3 \sim 0$). Moving closer to resonance causes an increased atom-light interaction, and the rotation rate increases. As the frequency becomes resonant with (in this case) the $\sigma^-$ transition (-4 GHz detuning), one hand of circular polarisation is absorbed completely, causing the polarisation vector to move to the pole ($S_1 = S_2 \sim 0$, $S_3 \sim 0.5$). Finally, as both transitions become resonant (-2 GHz detuning), all the light is absorbed and the polarisation vector disappears to the centre of the \Poincare sphere.
}
\label{fig:FaradayPoincare}
\end{center}
\end{figure*}
In this article, we introduce a visual understanding of magneto-optic rotation via the Faraday and Voigt effects based on rotation around an eigenaxis on the \Poincare sphere, and apply it to simple systems to demonstrate its use. 
Building on this understanding, we apply it to the problem of magneto-optic filters and show that by using a combination of Faraday and Voigt rotation in separate vapour cells, filter performance can be improved over what is possible by only using a single cell. 
Finally, we validate the concept with a quantitative comparison to experimental data, realising the highest performance Rb D2 line bandpass filter measured to date.

\section{Theory / Background}

The main theoretical framework for atom-light interactions in an atomic medium subject to an applied magnetic field has been covered extensively elsewhere~(see refs.\cite{Zentile2015b,Keaveney2017a} and references therein), but here we summarise some of the important points for understanding this work.

%
\begin{figure*}[t]
\begin{center}
\includegraphics[width=1.9\columnwidth]{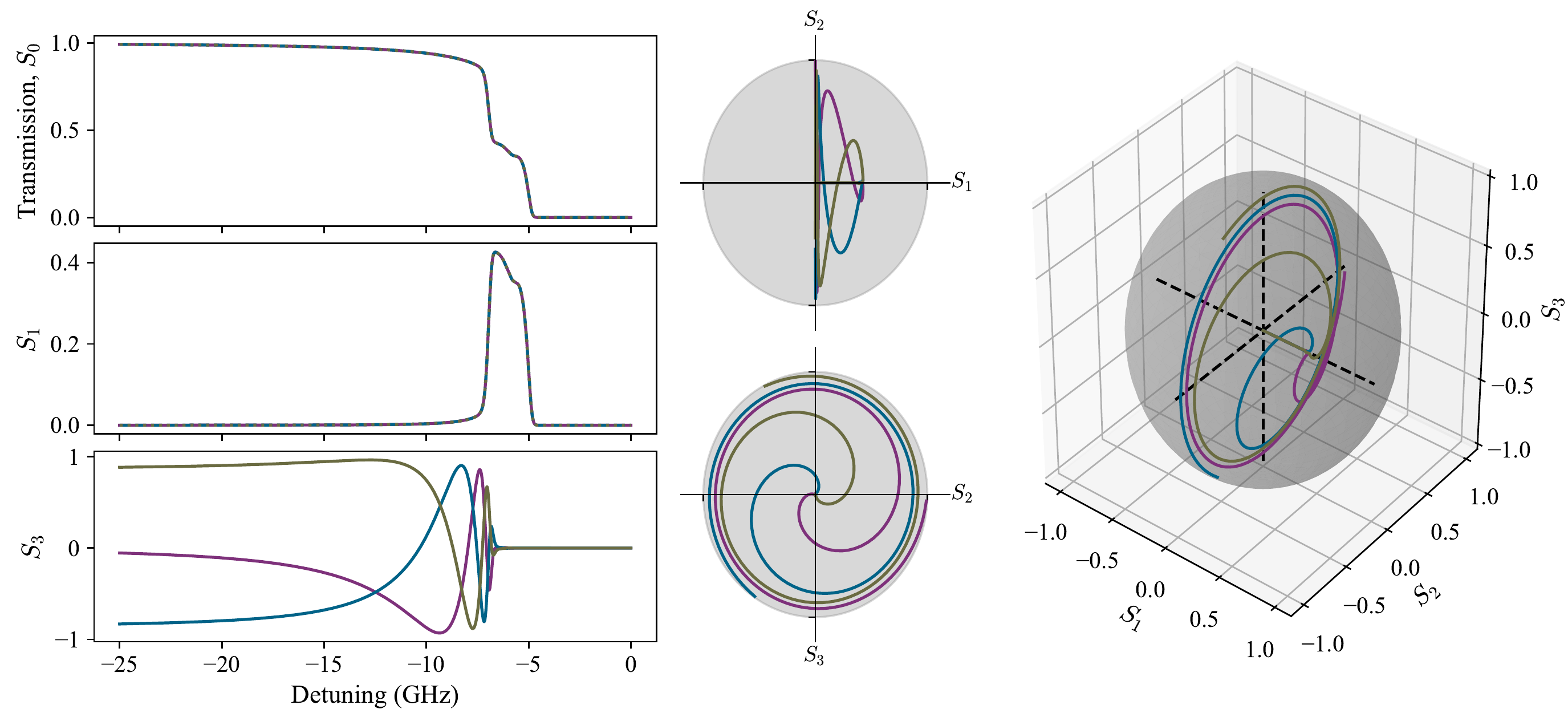}
\caption{Voigt rotation around the Rb D2 resonance lines visualised on the \Poincare sphere. This figure is an analog of figure~\ref{fig:FaradayPoincare} but for the Voigt geometry. We show trajectories for 3 initial polarizations $[E_x, E_y] = 1/\sqrt{2} [1, e^{i \phi}]$, $\phi = 0$ (purple), $2\pi/3$ (blue), $4\pi/3$ (olive) that differ in their relative phase between $x$ and $y$ components. In the same way as figure~\ref{fig:FaradayPoincare}, when viewed in the plane perpendicular to the rotation axis, the trajectories are self-similar. The difference between the two figures is the eigenbasis of light propagation and the coupling to atomic transitions. The far-from-resonance behaviour is much the same as the Faraday case, but in the Voigt geometry the polarisation rotates around the $S_1$ axis instead of the $S_3$ axis. 
Since the coupling to atomic transitions is different in the Voigt geometry, as the frequency becomes resonant with (in this case) the $\sigma^-$ transition (-6 GHz detuning), the component of linear polarisation perpendicular to the applied magnetic field vector is absorbed, and the polarisation vector therefore moves to $S_1 \sim 0.5$ before the $\pi$ transitions finally absorb the other component of polarisation.
}
\label{fig:VoigtPoincare}
\end{center}
\end{figure*}

\subsection{Stokes parameters and the \Poincare sphere}

The Stokes parameters provide a practical set of parameters to define the polarisation state of light, and involve measurements of the intensity in orthogonal bases. Since only intensities need to be measured, they are easily measurable with equipment commonly found in an optics laboratory. The four Stokes parameters are defined as
\begin{align}
S_0 &= \frac{I_x + I_y}{I_0} = \frac{I_L + I_R}{I_0}, \\
S_1 &= \frac{I_x - I_y}{I_0}, \\
S_2 &= \frac{I_\nearrow - I_\searrow}{I_0}, \\
S_3 &= \frac{I_L - I_R}{I_0},
\end{align}
where $I_0$ is the input intensity and the subscripts $x, y, \nearrow, \searrow, L$ and $R$ denote, respectively, the intensities measured after linear polarisers along the $x$ and $y$ axes, linear polarisers at $\pm \pi/4$ to the $x$ axis, and left/right circular polarisers.

The \Poincare sphere is a well-known and convenient way of visualising the polarisation state of light, where the three axes are $S_1, S_2$ and $S_3$. 
Though the Stokes parameters can be used to describe both polarised and unpolarised light fields, here we will only be concerned with fully polarised light. In this case, $S_0 = \sqrt{S_1^2 + S_2^2 + S_3^2}$ is represented by the radial coordinate on the \Poincare sphere (though note this is not true for unpolarised light).
Diametrically opposed points on the\Poincare sphere correspond to orthogonal polarisations, which will be an important point when discussing filtering applications in later sections.

%
\begin{figure*}[t]
\begin{center}
\includegraphics[width=1.7\columnwidth]{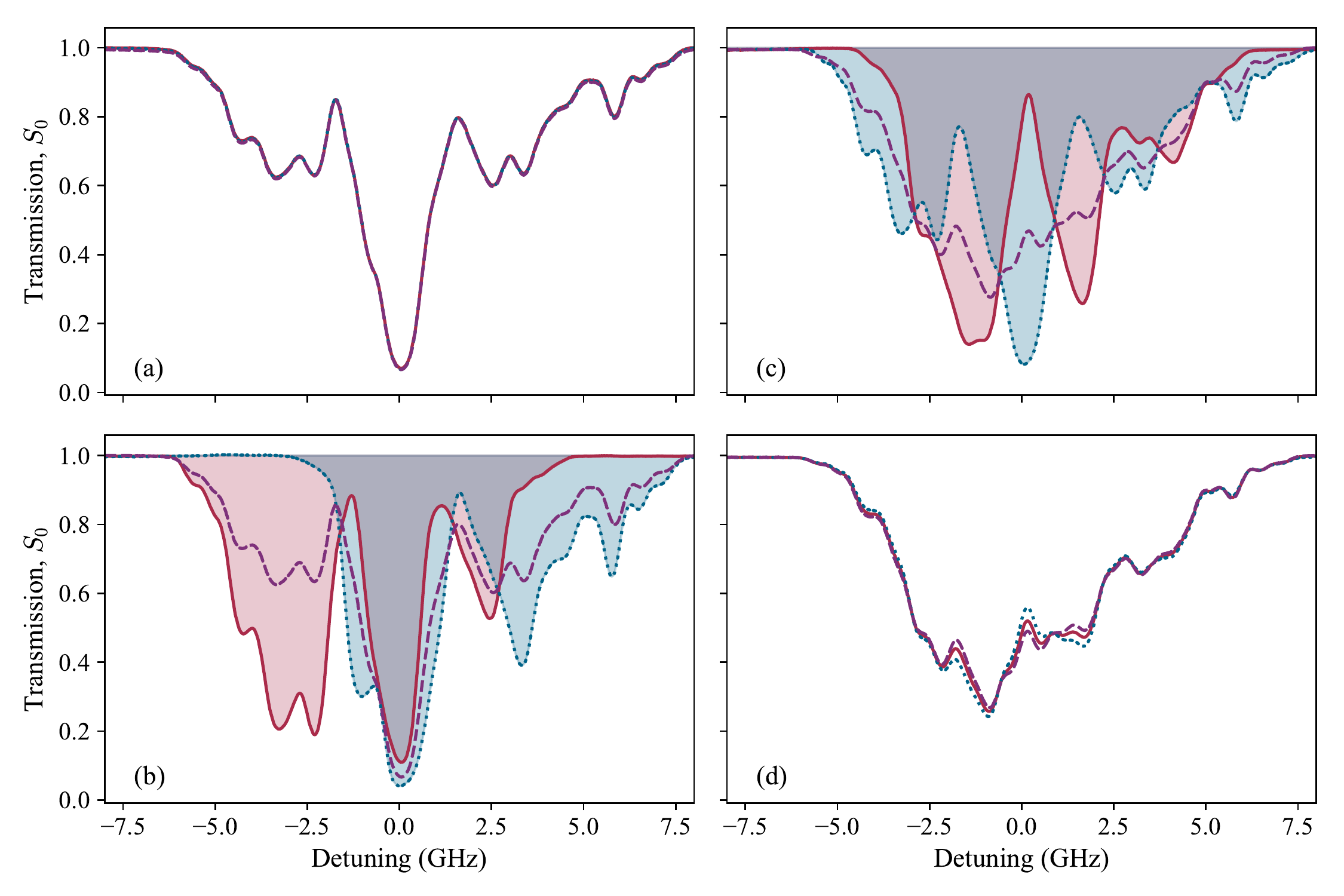}
\caption{Experimental transmission spectroscopy of the Rb D2 line in an applied field of 78~mT in the Faraday (a,b) and Voigt (c,d) geometries. Panels (a) and (c) show linearly polarised light with $\phi=0, \pi/4,\pi/2$ (red solid, purple dashed and blue dotted lines, respectively). Panels (b) and (d) show the two hands of circularly polarised light (red solid and blue dotted lines), and linearly polarised light with $\phi=\pi/4$ (purple dashed line). Shaded areas highlight the contributions to absorption in different regions of the spectrum from the various polarisation components.}
\label{fig:spec_data}
\end{center}
\end{figure*}

\subsection{Refractive indices and normal modes of propagation}

In order to calculate the coupling between the atomic vapour and the near-resonant light field that propagates through it, we require solutions to the wave equation,
\begin{align}
\vec{k} \times (\vec{k} \times \vec{E}) + \frac{\omega^2}{\epsilon_0 c^2} \epsilon \cdot \vec{E} = 0,
\end{align}
where $\vec{k}$ is the wavevector, $\vec{E}$ is the electric field, and $\omega$ is the angular frequency of the plane-wave. The dielectric tensor $\epsilon$ is related to the complex, frequency-dependent electric susceptibility $\chi(\omega)$ of the medium. The model we use to calculate $\chi(\omega)$ in the weak-probe~\cite{Sherlock2009} limit is called {\it ElecSus} and its operation is detailed in refs.~\cite{Zentile2015b,Keaveney2017a}.

Within the wave equation are the fundamentals of the interaction between light and a dielectric medium. 
We can simplify the wave equation with an appropriate choice of coordinate system. As discussed in refs.~\cite{Zentile2015b,Keaveney2017a} and references therein, the wave equation can then take a matrix form,
\begin{align}
\left(
\begin{array}{ccc}
( \epsilon_x - n^2) \cos(\theta_B) & \epsilon_{xy} & \epsilon_x \sin(\theta_B) \\
-\epsilon_{xy} \cos(\theta_B)  & \epsilon_x - n^2 & -\epsilon_{xy} \sin(\theta_B) \\
(n^2 - \epsilon_z) \sin(\theta_B) & 0 & \epsilon_z \cos(\theta_B)
\end{array}
\right)
\left(
\begin{array}{ccc}
E_x \\ E_y \\ E_z
\end{array}
\right)
= 0,
\label{eq:wavematrix}
\end{align}
where $n$ is a refractive index of the medium and $\theta_B$ is the angle the applied magnetic field makes with the light propagation axis $\vec{k}$.
The dielectric tensor elements $\epsilon_x$, $\epsilon_{xy}$, and $\epsilon_z$ are related to elements of the electric susceptibility $\chi_0,\; \chi_-$ and $\chi_+$ (associated with $\pi$, $\sigma^-$, and $\sigma^+$ transitions, respectively) by
\begin{align}
\epsilon_x &= \frac{1}{2} ( 2 + \chi_+ + \chi_- ), \\
\epsilon_{xy} &= \frac{\rm i}{2} (\chi_- - \chi_+ ), \\
\epsilon_z &= 1 + \chi_0.
\end{align}
Solutions for this set of equations are found by setting the determinant of the matrix to zero, which results in a quadratic equation in $n^2$. 
The two solutions for $n$ are used to find the normal modes for the system, and we find the system has two refractive indices that couple in different ways to components of the light field - in other words, the system is generally birefringent and dichroic.
Whlist this system can be solved for any arbitrary angle of $\theta_B$, the solutions must be numerically computed and much of the physical insight is lost. Magneto-optic effects with a non-axial, non-perpendicular magnetic field are discussed relatively rarely in the literature~\cite{Menders1992,Edwards1995,Nienhuis1998,Rotondaro2015,Keaveney2018a}. 
In this work we focus on the two special cases where $\theta_B=0$ or $\theta_B = \pi/2$. These two angles form the Faraday and Voigt geometries, respectively, and for clarity we show these arrangements in figure~\ref{fig:geometry}.

%
\begin{figure}[t]
\begin{center}
\includegraphics[width=\columnwidth]{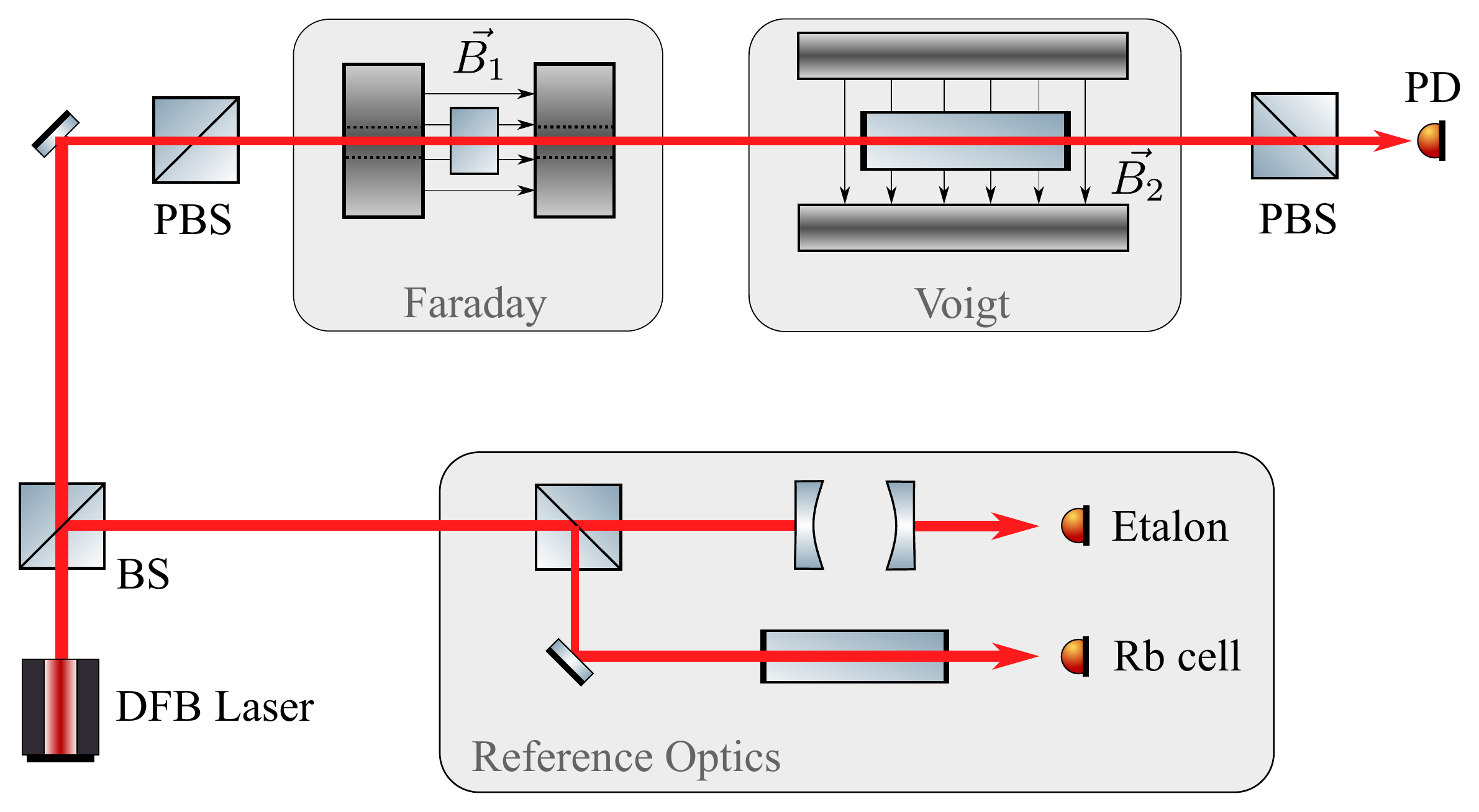}
\caption{Schematic of the experimental setup. A DFB laser at 780~nm is scanned across the Rb D2 line. The filter comprises crossed-polarisers (PBS), between which are placed two vapour cells. The first vapour cell is 5~mm long and placed in an axial magnetic field $\vec{B}_1$ formed by two cylindrical permanent magnets, while the second cell  is 50~mm long and placed in a transverse field $\vec{B}_2$. Reference optics made up of a room temperature zero-field Rb vapour cell and a Fabry-Perot etalon allow the laser scan to be calibrated.
}
\label{fig:ExptSetup}
\end{center}
\end{figure}

The Faraday geometry is most commonly used for atomic bandpass filters, and is easy to understand both conceptually and mathematically. In this geometry, the solutions for the two refractive indices of the medium are~\cite{Keaveney2017a}
\begin{align}
n_1 &= \sqrt{\epsilon_x - {\rm i}\epsilon_{xy}} = \sqrt{1 + \chi_+} \\
n_2 &= \sqrt{\epsilon_x + {\rm i}\epsilon_{xy}} = \sqrt{1 + \chi_-},
\end{align}
or alternately, that the two indices are associated with $\sigma^+$ and $\sigma^-$ transitions. 
The corresponding eigenvectors are
$\vec{e}_1 = (i,1,0)^T$ and $\vec{e}_2 = (-i,1,0)^T$.
Applying either of the two eigenvectors simply transforms a set of coordinates in the cartesian basis into a circular basis, and hence we find the well-known solutions that the two circular polarisation components couple directly and independently to the $\sigma^\pm$ atomic transitions. The exact coupling of left/right handedness to $\pm$ transition depends on the exact orientation of the magnetic field (i.e. parallel or anti-parallel). The atomic $\pi$ transitions are never driven in this geometry. Since $n_1$ and $n_2$ are, in general, different, the system is birefringent (and dichroic, since $n_{1,2}$ are complex), and because they couple to the circular polarisation states of light, the system exhibits {\it circular} birefringence. On propagation through a medium, a phase difference between circular polarisation components is generated, which when mapped back to the linear basis is viewed as a rotation of the plane of linear polarisation -- the well-known Faraday effect~\cite{Faraday1846}.

For the Voigt geometry, we have a different set of solutions for $n_1$ and $n_2$, which are
\begin{align}
n_1 &= \sqrt{\epsilon_x + \epsilon_{xy}^2/\epsilon_x} = \sqrt{\frac{2(1+\chi_+ + \chi_- + \chi_+ \chi_-)}{(2+\chi_+ + \chi_-)}} \\
n_2 &= \sqrt{\epsilon_z} = \sqrt{1 + \chi_0}.
\end{align}
We find that $n_1$ is associated with both $\sigma^\pm$ transitions, while $n_2$ is associated with only $\pi$ transitions. 
The normal modes, for $\theta_B = \pi/2$ (i.e. $\vec{B} = \vert B \vert \hat{x}$), are 
$\vec{e}_1 = (0,\epsilon_x / \epsilon_{xy}, 1)^T$ and $\vec{e}_2 = (1,0,0)^T$.
We see immediately from $\vec{e}_2$ that the electric field component of the light that is parallel to the magnetic field (along $x$) couples to $\pi$ transitions through $n_2$. The first eigenmode $\vec{e}_1$ is elliptically polarised in the plane perpendicular to $\vec{B}$, i.e. in the $yz$-plane, and couples to both $\sigma^+$ and $\sigma^-$ transitions through $n_1$. This mode is an electric field polarised perpendicular to the magnetic field axis (i.e. along $y$). The two eigenmodes are therefore the orthogonal linear polarisations, and the system exhibits {\it linear} birefringence and dichroism. As for the Faraday effect, a phase change develops on propagation through a medium and there is a polarisation rotation known as the Voigt effect~\cite{Voigt1898}.

In both the Faraday and Voigt geometries, the two eigenmodes are orthogonal and therefore form an eigenaxis on the \Poincare sphere. In the next section we will see how we can use this in a visual understanding of magneto-optic rotation.

%
\begin{figure*}[t]
\begin{center}
\includegraphics[width=1.65\columnwidth]{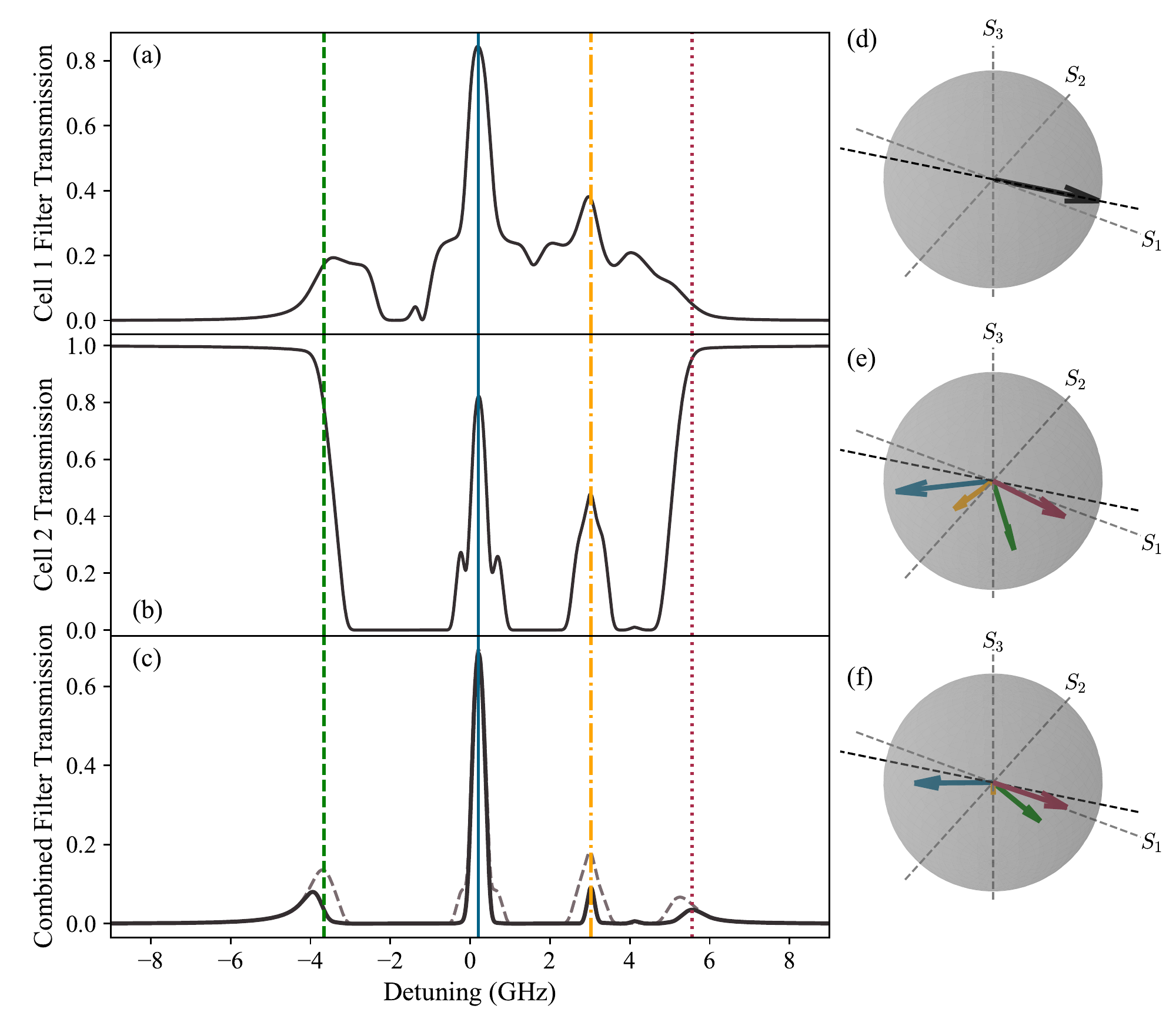}
\caption{Construction of a two-cell filter system. Panel (a) shows the filter profile if cell 2 is absent, while panel (b) shows the transmitted spectrum through the second cell with a frequency-dependent input polarisation based on the output from cell 1. Panel (c) shows the final dual-cell filter profile, which utilises both the extra absorption and optical rotation caused by the second cell. For comparison, the dashed grey line in panel (c) shows the expected filter profile if there was only absorption and no rotation from the second cell. Panel (d) shows the (frequency-independent) input polarisation vector on the \Poincare sphere; the dashed black line in (d-f) shows the axis of this polarisation vector - the ideal filter will have an output vector pointing in the opposite direction to the input vector. Panels (e) and (f) show the output polarisation at selected frequencies (marked by vertical lines in panels (a-c)) after cell 1 and after cell 2, respectively, which highlight the role of extra optical rotation caused by the second cell.}
\label{fig:construction}
\end{center}
\end{figure*}

\section{Visualising magneto-optic rotation on the \Poincare sphere}

Using {\it ElecSus}~\cite{Zentile2015b,Keaveney2017a}, we calculate the Stokes parameters as a function of frequency detuning for the Rb D2 line. Note that detuning is defined with respect to the weighted line-center of the D2 line~\cite{Siddons2008b}. Figure~\ref{fig:FaradayPoincare} shows an example spectrum in the Faraday geometry with a magnetic field strength of 50~mT. This magnetic field is large enough to split the resonance lines by more than the Doppler-broadened line width, so there are areas of the spectrum where one transition dominates the absorption.

On the left side of figure~\ref{fig:FaradayPoincare} we show $S_0$, $S_1$ and $S_3$ spectra on the red-detuned side of resonance. We plot the output Stokes parameters for 3 different input polarisations. All three are linearly polarised, with $\pi/3$ radians difference between them. For the $S_0$ and $S_3$ spectra, the initial input polarisation does not affect the output, but for the $S_1$ spectrum, the spectra are different. All three show the characteristic oscillation that is due to Faraday rotation, but it is not immediately obvious that these spectra share any other common features. Once we look at their trajectories along the \Poincare sphere, however, the relationship between the spectra becomes more clear. The $\pi/3$ difference in polarisation angle maps into a three-fold rotational symmetry of their trajectory on the \Poincare sphere; other than their starting coordinates in the $S_1,S_2$ plane, the trajectories are identical - we call this self-similarity.

The trajectory around the \Poincare sphere can be understood as follows. Far from resonance, the polarisation vector starts at the equator of the \Poincare sphere in the $S_1,S_2$ plane. As the detuning becomes more positive, the vector rotates slowly around the equator owing to a small difference in the refractive indices $n_1$ and $n_2$. At this point the absorption is almost negligibly small, so the optical rotation still maintains linear polarisation (i.e. $S_3 \sim 0$). Eventually, as the frequency is increased further towards resonance, the atom-light interaction becomes stronger, which causes an increased rotation rate, but at the same time, absorption of one component of circular polarisation becomes significant. The transmitted light is therefore predominantly circular (i.e. the component which is not absorbed), and $\vert S_1 \vert \rightarrow 0$ as $S_3 \rightarrow \pm0.5$ (the sign depends on which hand of light is absorbed).
Finally, as both transitions become resonant (around -2 GHz detuning in this case), all the light is absorbed ($S_0 \rightarrow 0$) and the polarisation vector disappears to the centre of the \Poincare sphere.

To look at the difference the magnetic field direction makes we compare figure~\ref{fig:FaradayPoincare} with figure~\ref{fig:VoigtPoincare}, which is an example of optical rotation in the Voigt geometry. In this case we have an transverse applied magnetic field strength of 100~mT.
We have chosen the parameters of the two systems deliberately in order to draw parallels between the two systems. The similarities between the two figures are clear, except that circular birefringence is swapped for linear birefringence. In the \Poincare picture, this changes the axis of optical rotation; since the eigenmodes of the system change from the $L,R$ circular polarsiations to the $x,y$ polarisations, the eigenaxis of rotation swaps from the $S_3$ axis to the $S_1$ axis.

%
%
\begin{figure}[t]
\begin{center}
\includegraphics[width=0.9\columnwidth]{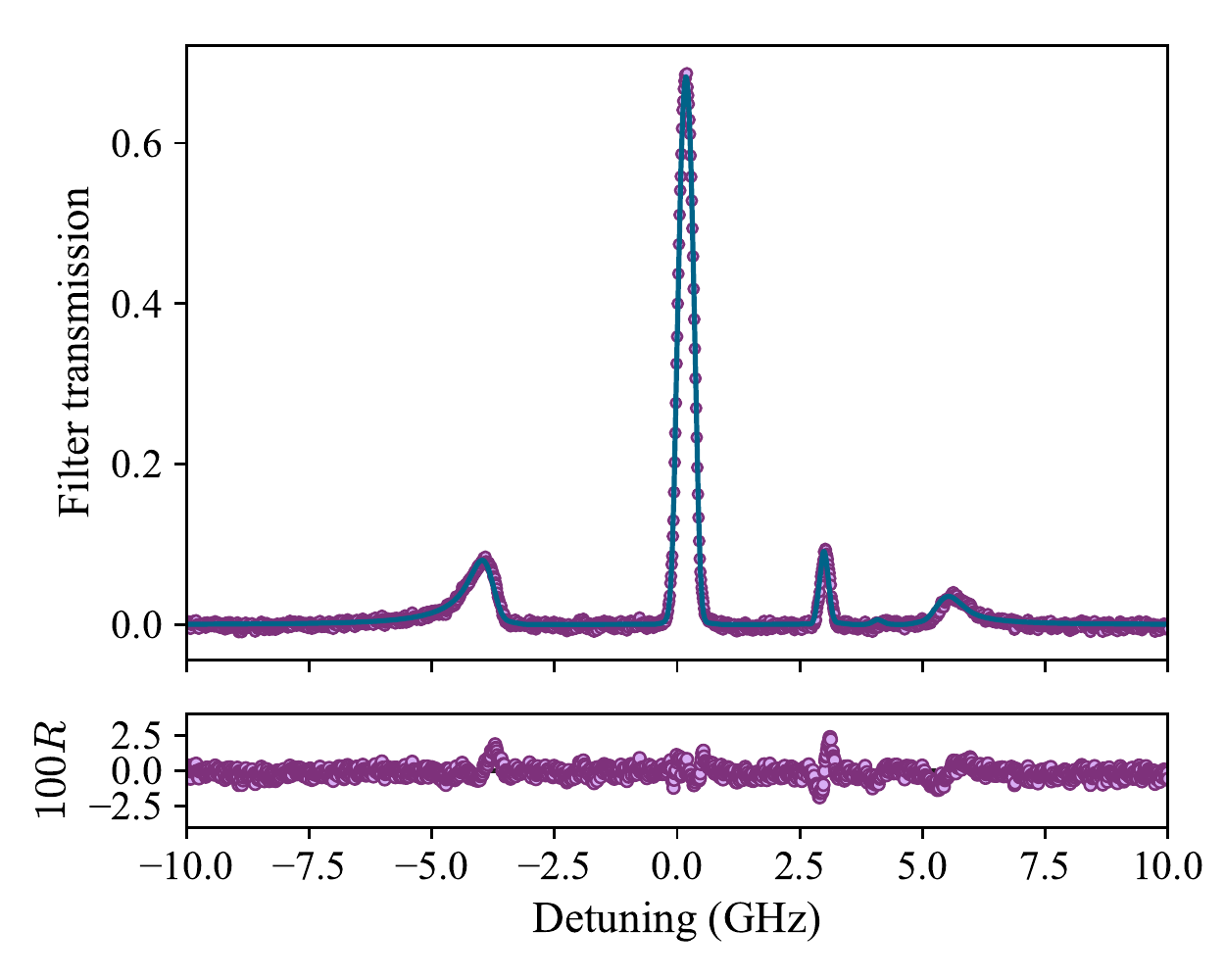}
\caption{Experimental demonstration of a dual-cell optical filter. Purple points are experimental data, while the blue line is a fit to the data using the model described in the text. The RMS error between theory and experiment is 0.4\%, and the residuals $R$ are small and almost structureless, indicating an excellent fit~\cite{Hughes2010}.}
\label{fig:ExptDemo}
\end{center}
\end{figure}
%

\section{Spectroscopy in the Faraday and Voigt geometries}

Figure~\ref{fig:spec_data} shows experimental transmission ($S_0$) spectra for the Rb D2 line in a 5~mm long naturally abundant Rb vapour cell. The cell is placed in a magnetic field strength of 78~mT in both Faraday and Voigt configurations, with a variety of input polarisations. Panels (a) and (b) show data in the Faraday geometry, while panels (c) and (d) show data in the Voigt geometry ($\vec{B} = \vert B \vert \hat{x}$). We plot linear polarisation aligned along $x$ (red), along $y$ (blue) and at $\pi/4$ to the $x$-axis (purple) in panels (a) and (c). In panels (b) and (d) we plot the two circular polarisations (red/blue) and linear polarisation at an angle $\pi/4$ to the $x$ axis (purple).
For the Faraday geometry, the angle of linear polarisation makes no difference to the spectra, since any linear polarisation is an equal mixture of the two components in the circular basis. Swapping between left and right circular polarisation in the Faraday geometry changes which of the $\sigma^\pm$ transitions are driven, and makes a considerable difference to the spectra, as can be seen in panel (c) where, for clarity, we have also filled under the areas in red and blue. Though the magnetic field is not high enough to completely separate the transitions, the edges of the two spectra are predominantly driven by $\sigma^-$ transitions at negative detuning, and $\sigma^+$ at positive detuning.

For the Voigt geometry, again we have the opposite behaviour to the Faraday case. In panel (b), the two orthogonal linear polarisation components $x$ and $y$ drive $\pi$ and $\sigma^\pm$ transitions, respectively. We have again highlighted the two types of transition with filled areas. In this case there is no clear structure, since the Zeeman splitting from the magnetic field is not large enough to dominate over the internal hyperfine structure of Rb. Finally, panel (d) shows that in the Voigt geometry, the transmission spectra for left and right circular polarisation are the same as for linear polarisation at an angle of 45 degrees to the magnetic field axis - both $\pi$ and $\sigma^\pm$ transitions are driven equally by all three input polarisations.

\section{Application to optical bandpass filters}

We now apply our understanding of magneto-optic rotation to optical filtering. In order to quantify our results, we first define some performance metrics for optical filters.
Broadly speaking, one desires an optical filter to have a high transmission at the frequencies of interest, while having a high out-of-band extinction. A narrower transmission peak in frequency allows for a more spectrally selective filter, with better noise rejection, and is therefore a desirable feature.
While a full-width at half maximum could be used, this is a bad measure if multiple transmission peaks are present in the spectrum (which is often the case with atomic filters). A better approach is to use the equivalent noise bandwidth (ENBW), which takes into account the out-of-band transmission of the filter at all frequencies.
The ENBW is defined as
$\mrm{ENBW}=\frac{\int \mathcal{T}(\omega)\mrm{d}\omega}{\mathcal{T}(\omega_s)}$,
where $\cal{T}$ is the transmission of the filter, $\omega$ is the angular optical frequency and $\omega_s$ is the signal angular frequency. For most situations, the signal frequency coincides with the peak filter transmission, but can be set differently if specific frequencies are required. 
While ENBW could be used alone, this is often not the best performance metric, since the narrowest ENBW will also coincide with a filter which blocks all frequencies. Since a high peak transmission with a narrow bandwidth is clearly a desirable feature, a better figure-of-merit (FOM) is to use the ENBW scaled by the peak filter transmission, such that 
$\mrm{FOM} = {\mathcal{T}(\omega_s)}/{\mrm{ENBW}}$
This approach was first suggested by ref.~\cite{Kiefer2014}. Using this FOM, we can maintain high transmission while minimizing ENBW. 
This FOM has previously been used to find the optimal performance of atomic Faraday filters~\cite{Zentile2015c,Zentile2015a}, and in the case of an unconstrained magnetic field angle to theoretically optimize, and experimentally demonstrate, improved performance on the Rb D2 line~\cite{Keaveney2018a}.


The idea of using multiple cells for improved filter performance has been suggested in the past, notably in refs.~\cite{Cimino1968,Yeh1982}, but used the second cell only as an absorption medium to reduce some of the out-of-band transmission. Another recent implementation of a two-cell filter combined a Faraday filter with a stimulated Raman gain amplifier~\cite{Zhao2015}, but this requires an additional laser system to pump the medium.
Our approach is novel in that we combine the optical rotation from two cells with different conditions. To the best of our knowledge this has not been done before. 
We first use a Faraday geometry cell and then a Voigt geometry cell to take advantage of the two orthogonal optical rotation axes discussed in the previous sections. 
Since the rotation axes are orthogonal, we can take advantage of the self-similarity property to gain additional freedom between the two cells. 
We start with linearly polarised light, but are free to choose the angle of the input polariser $\phi$ with no consequence to the output spectrum after the Faraday cell (recall figure~\ref{fig:FaradayPoincare}), provided that the output polariser is also rotated to maintain crossed-polarisers.
The input polarisation angle does, however, affect the propagation through the Voigt cell, so we gain an essentially independent parameter which we can use to improve filter performance.
We have performed a computer optimisation routine to find the best filter performance, with some constraints based on experimentally available cells and conditions in order to later demonstrate the filter experimentally. 
The cell lengths are fixed (5~mm in the Faraday cell, 50~mm in the Voigt cell), as are the isotopic abundances (natural abundance ratio of Rb) and magnetic field directions, and we confine our investigation to the Rb D2 line. 
We optimise five parameters of the system - for each cell, we can vary the magnetic field strength $\vert B \vert$ and cell temperature $T$, which sets the atomic number density and therefore the overall strength of the atom-light interaction. 
We also vary the initial angle of linear polarisation $\phi$, which, as already discussed, affects the coupling to atomic transitions in the Voigt cell. For the Faraday cell, our starting parameters were the same as the intra-cavity filter used in ref.~\cite{Keaveney2016c}.
The starting parameters for the Voigt cell were chosen empirically. 
Since this kind of optimisation problem is likely to have many local minima/maxima in parameter space, we do not claim this is the best global solution, but it serves as a proof-of-principle demonstration.

The operating principle of the optimised (with constraints) two-cell filter is shown in figure~\ref{fig:construction}.
Fig.~\ref{fig:construction}(a) shows what the output of the filter would be without the Voigt cell; fig.~\ref{fig:construction}(b) shows the relative transmission through the Voigt cell, with a frequency-dependent input polarisation that comes from the output polarisation of the Faraday cell. 
Finally, fig.~\ref{fig:construction}(c) shows the dual-cell optical filter spectrum in a solid black line. 
We also plot a dashed grey line, which is the result if we neglect the extra optical rotation from the Voigt cell and instead assume that this cell acts only as an absorption medium. 
It is clear from examining these two curves that the extra rotation makes a large difference to the output spectrum. 
This is further highlighted by the \Poincare sphere representations - panel (d) shows the frequency-independent input polarisation vector, which is at a slight angle $\phi \approx 6^\circ$ with respect to the $x$-axis. Panels (e) and (f) show the output after the first and second cells at four frequencies that are shown as colour-coded vertical lines in panels (a-c).
In the transmission band (blue, solid line in (a-c)), the Faraday cell rotates the plane of linear polarisation as near as possible to lie close to an eigenaxis in the Voigt cell. Since the incoming light in the Voigt cell is close to an eigenaxis, there is little further rotation at this frequency and the transmitted peak is only partially attenuated. For the green (dashed) and yellow (dot-dashed) frequencies, a comparison of panels (e) and (f) shows that there is still significant optical rotation in addition to attenuation. The polarisation at the frequency of the final small peak (red, dotted line) after the Faraday cell lies very close to the eigen-axis of the Voigt cell and so is not significantly rotated- but this also lies close to the rejection axis of the second polariser and so the transmission at this frequency is very small.

\subsection{Experimental verification}

In order to test the optimisation result, we have performed an experiment to quantitatively demonstrate the filter performance.
The experimental arrangement is shown in figure~\ref{fig:ExptSetup}.
The filter is comprised of two polarisers (PBS cubes placed at right-angles to each other forming crossed-polarisers) between which we place two vapour cells and two sets of magnets. The first cell is 5~mm long, filled with Rb in its natural abundance ratio and placed between two cylindrical permanent (NdFeB) magnets that create an axial magnetic field with adjustable strength via the magnets' separation. At the closest position of the two magnets, the field strength is around 0.5~T. The field variation across the cell depends on the separation of the two magnets; for our arrangement (26.5~mT field strength) the variation is calculated to be less than 1\%.
The second cell is 50~mm long, filled with Rb in its natural abundance ratio and is placed between two cuboid-shaped magnets. The plate-magnets have dimensions of $150\times 50 \times 10$ mm, are placed with their long axis along the light propagation direction and are magnetised along the short axis in order to create a transverse magnetic field over the vapour cell. Adjusting the separation alters the field strength in the cell; we operate at a field strength of 22~mT which requires a separation of 18~cm.
The field variation over the length of the cell, calculated from a finite-element model (we use the Radia plug-in for Mathematica) is less than 2\% of the average field strength.
The two cells are placed sufficiently far apart that the magnetic fields experienced by each cell are independent. The vapour cells can be heated independently to provide the required atomic number density. 

A weak-probe~\cite{Sherlock2009} beam is used to determine the filter transmission. We use around 100 nW of optical power, focussed in the 5~mm vapour cell to a 1/e$^2$ waist of approximately 100~$\mu$m (lenses not shown in schematic), and collimated with a width of approximately 2~mm in the 50~mm cell. The output after the second polariser is detected on a photodiode and the 780~nm DFB laser is scanned across the Rb D2 resonance lines. To calibrate the laser frequency scan, reference optics are used, following the method in ref.~\cite{Keaveney2014a}. The filter transmission is normalised by taking a reference transmission level, obtained by rotating the output polariser so that the transmission is maximised.

Figure.~\ref{fig:ExptDemo} shows the experimental filter spectrum (purple data points). To account for experimental uncertainties, we perform a least-squares fit~\cite{Hughes2010} to the data with the same model used to predict the optimal parameters. The fit and experimental data are in excellent agreement, with a RMS error between experiment and theory of 0.4\% and almost structureless residuals~\cite{Hughes2010}. The fit parameters, where the subscripts $1,2$ refer to the Faraday and Voigt cells, respectively, are $T_{\rm 1} = 86.7\pm0.3^\circ$C, $B_{\rm 1} = 27\pm1$~mT, $T_{\rm 2} = 79.0\pm0.3^\circ$C, $ B_{\rm 2} = 24\pm4$~mT, and $\phi = 6\pm1^\circ$ are within 1\% (B, T) and 1 degree of the theoretical optimum parameters.

The experimental FOM value is $1.22\pm0.02$~GHz$^{-1}$, which matches with the theoretically predicted value of 1.24~GHz$^{-1}$. The central peak has a FWHM of 350~MHz, the ENBW of the filter is 0.56~GHz, and the peak transmission is 68\%. This filter has the highest FOM measured on the Rb D2 line to date.

\section{Conclusion}

In conclusion, we have demonstrated a detailed understanding of magneto-optic rotation in the Faraday and Voigt geometries, and illustrated this with a graphical method based on trajectories around the \Poincare sphere. Applying this to optical bandpass filters, we have designed and optimised, subject to experimental constraints, a filter that uses the optical rotation from two cells in tandem. This dual-cell filter has a better performance FOM than any other Rb D2 line filter measured to date. Future improvements could be made by combining the dual-cell filter with an unconstrained magnetic field geometry~\cite{Keaveney2018a} or additional polarisation optics between the two cells.
The data presented in this paper are available from DRO\footnote{\url{DOI://xxx.xxxxxxx}, added at proof stage}.

{\bf {Funding}} EPSRC (Grant No. EP/R002061/1) and Durham University. 

\small
\bibliography{library}

\end{document}